\documentclass{elsart}
\usepackage[square,comma]{natbib}
\usepackage{graphicx}
\usepackage{amssymb}

  \def\bld#1{{\bf #1 }}

\def\part#1#2{\frac{\partial #1}{\partial #2}}

\def\rb{\right)}
\def\lb{\left(}

\begin{document}

\begin{frontmatter}


\title{CHALLENGES IN NON LINEAR GRAVITATIONAL CLUSTERING }

\author{T. Padmanabhan}

\address{Inter-University Centre for Astronomy and Astrophysics,\\
 Post Bag 4, Ganeshkhind, \\
Pune-411 007, India.}
\ead{nabhan@iucaa.ernet.in}

\begin{abstract}
This  article addresses some issues related to statistical description of
gravitating systems  in  an expanding backgrounds. In particular, I describe 
(a) how the non linear mode-mode coupling transfers power from one scale to another in the Fourier space if the initial power spectrum is sharply peaked at a given scale and
(b) what are the asymptotic characteristics  of gravitational clustering
that are independent of the initial conditions. The analysis uses a new approach based on 
an integro-differential equation for the evolution of the gravitational potential in the Fourier space. I show how this equation allows one to understand several aspects of nonlinear gravitational clustering and provides insight in to the transfer of power from one scale to another through nonlinear mode coupling. Numerical simulations as well as analytic work shows that  power transfer leads to a universal power spectrum at late times, somewhat reminiscent of the existence of Kolmogorov spectrum in fluid turbulence.

\end{abstract}


\end{frontmatter}

\section{\label{intro}Introduction }
One of the central problems in cosmology is to describe
  the non linear phases of the gravitational clustering starting from an initial spectrum of density
  fluctuations. It is often enough (and necessary) to use a statistical description and  relate
  different statistical indicators (like the  power spectra, $n$th order  correlation functions etc.)
  of the resulting density distribution to the statistical parameters (usually the power spectrum) of the 
  initial distribution. The relevant scales at which gravitational clustering is non linear are less than
  about 10 Mpc (where 1 Mpc = $3\times 10^{24}$ cm is the typical separation between galaxies in the
  universe) while the expansion of the universe has a characteristic scale of about few thousand 
  Mpc. Hence  non linear gravitational clustering in an expanding universe can 
  be adequately described by Newtonian gravity provided the rescaling of lengths due to the
  background expansion
  is taken into account. 
  
 As to be expected, cosmological expansion  introduces  
  several new factors into the problem when compared with the study of statistical mechanics of isolated
  gravitating systems. (For a general review of statistical mechanics of gravitating systems,
  see \cite{tppr}. For a sample of different approaches, see \cite{sample} and the references cited therein. 
  Review of gravitational clustering in expanding background is also available in several textbooks in cosmology \cite{cosmotext,lssu}.)
  Though this problem can be
tackled in a `practical' manner using high resolution numerical simulations (for a review, see \cite{numsim}),
such an approach hides the physical principles which govern 
the behaviour of the system. To understand the physics, it is necessary to
attack the problem from several directions using analytic and semi analytic
methods. Several such attempts exist in the literature based on Zeldovich(like) approximations \cite{za},
path integral and perturbative techniques \cite{pi}, nonlinear scaling relations \cite{nsr} and many others. In spite of all these it is probably fair to say that we still do not have a clear analytic grasp of this problem, mainly because each of these approximations have different domains of validity and do not arise from a central paradigm.

I propose to attack the problem from a different angle, which has not received much attention in the past. The approach begins from the dynamical equation for the the density contrast in the Fourier space  and casts it as an integro-differential equation. Though this equation is known in the literature  (see, e.g. \cite{lssu}), it has received very little attention  because it is not `closed' mathematically; that is, it involves variables which are not natural to the formalism and thus further progress is difficult. I will, however, argue that
there exists a natural closure condition for this
equation  based on Zeldovich approximation thereby allowing us to write down
 a \textit{closed} integro-differential equation for the gravitational potential
in the Fourier space. 

It turns out that this equation can form the basis for several further investigations some of which are described in ref. \cite{gc1} and in the second reference in \cite{tppr}. Here I will concentrate on just two specific features, centered around the following issues:

\begin{itemize}
\item
   If the initial power spectrum is sharply peaked in a narrow band of wavelengths, how does
  the evolution transfer the power to other scales? In particular, does the non linear evolution in the case
of gravitational interactions lead to a universal power spectrum (like the Kolmogorov spectrum in fluid turbulence)? 
 
  \item
  What is the nature of the time evolution at late stages? Does the gravitational clustering at late stages wipe out the memory of initial conditions and evolve in a universal manner?
\end{itemize}

\section{\label{gravclnl}An integral equation to describe nonlinear gravitational clustering}

The expansion of the universe sets a natural length scale (called the Hubble radius) $d_H = c
(\dot a/a)^{-1}$ which is about 4000 Mpc in the current universe. 
In any region which is small compared to $d_{\rm H}$ one can set up an unambiguous coordinate system in which the {\it proper} coordinate of a particle ${\bf r} (t)=a(t){\bf x}(t)$ satisfies the Newtonian equation $\ddot {\bf r} = -  {\nabla }_{\bf r}\Phi$ where $\Phi$ is the gravitational potential. 
The Lagrangian for such a system of particles is given by
\begin{equation}
L=\sum_i\left[
\frac{1}{2}m_i\dot{\mathbf{r}}_i^2+\frac{G}{2}\sum_j\frac{m_im_j}{|\mathbf{r}_i-\mathbf{r}_j|}
\right]
\end{equation} 
In the term
\begin{equation}
\frac{1}{2}\dot{\mathbf{r}}_i^2
=\frac{1}{2}\left[a^2\dot{\mathbf{x}}_i^2+\dot{a}^2\mathbf{x}_i^2+a\dot{a}\frac{d\mathbf{x}_i^2}{dt}\right]
=\frac{1}{2}\left[a^2\dot{\mathbf{x}}_i^2-a\ddot{a}\mathbf{x}_i^2 +\frac{da\dot{a}\mathbf{x}_i^2}{dt}\right]
\end{equation} 
we note that: (i) the total time derivative can be ignored; (ii) using $\ddot{a}=-(4\pi G/3)\rho_ba$, the term $\Phi_{FRW}=-(1/2)a\ddot{a}\mathbf{x}_i^2
=(2\pi G / 3)\rho_0 (x_i^2/a) $ can be identified as the gravitational potential due to the uniform Friedman background
of density $\rho_b=\rho_0/a^3$. Hence the Lagrangian can be expressed as
\begin{equation}
L=\sum_i m_i\left[\frac{1}{2}a^2\dot\mathbf{x}_i^2 +\phi(t,\mathbf{x}_i)\right]=T-U
\label{basicL}
\end{equation}
where
\begin{equation}
\phi=-\frac{G}{2a}\sum_{j}\frac{m_j}{|\mathbf{x}_i-\mathbf{x}_j|}-\frac{2\pi G\rho_0}{3}\frac{x_i^2}{a}
\label{defphi}
\end{equation} 
 is the difference between the total potential and the potential for the back ground Friedman universe $\Phi_{FRW}$. Varying the Lagrangian in Eq.(\ref{basicL}) with respect to $\mathbf{x}_i$, we get the equation of motion to be:
\begin{equation} 
\ddot{\bf x} + 2 {\dot a \over a}\dot{\bf x} = - {1 \over a^2} \nabla_x \phi\ 
\label{traj}
\end{equation}
Since $\phi$ is the gravitational potential generated by the \textit{perturbed}  mass density, 
  it satisfies the equation with the source $(\rho-\rho_b)\equiv
 \rho_b\delta$:
\begin{equation}
 \nabla^2_x \phi  = 4 \pi G \rho_ba^2 \delta 
 \label{poisson}
  \end{equation}
 Equation~(\ref{traj}) and Eq.(\ref{poisson}) govern the nonlinear gravitational clustering in an expanding background.

Usually one is interested in the evolution of the density contrast $\delta \lb t, \bld x \rb $ rather than in the trajectories. Since the density contrast can be expressed in terms of the trajectories of the particles, it should be possible to write down a differential equation for $\delta (t, \bld x)$ based on the equations for the trajectories $\bld x (t)$ derived above. It is, however, somewhat easier to write down an equation for $\delta_{\bld k} (t)$ which is the spatial Fourier transform of $\delta (t, \bld x)$. To do this, we begin with the fact that the density $\rho(\bld x,t)$ due to a set of point particles, each of mass $m$, is given by
\begin{equation}
\rho (\bld x,t) = {m\over a^3 (t)} \sum\limits_i \delta_D [ \bld x - \bld x _i(t)]
\end{equation}
where $\bld x_{i}(t)$ is the trajectory of the ith particle and $\delta_D$ is the Dirac delta function.  The density contrast $\delta (\bld x,t)$ is related to $\rho(\bld x, t)$ by
\begin{equation}
1+\delta (\bld x,t) \equiv {\rho(\bld x, t) \over \rho_b} = {V \over N} \sum\limits_i \delta_D [\bld x - \bld x_i(t)] =  \int d {\bld q} \delta_D [\bld x - \bld x_{T} (t, \bld q)]  . 
\end{equation}
In arriving at the last equality we have taken the continuum limit by: (i) replacing $\bld x_i(t)$ by $\bld x_T(t,\bld q)$ where  $\bld q$ stands for a set of parameters (like the initial position, velocity etc.) of a particle; for
simplicity, we shall take  this to be initial position. The subscript `T' is just to remind ourselves that $\bld x_T(t,\bld q)$ is the
\textit{trajectory} of the particle. (ii) replacing $(V/N)$ by $d^3{\bld q}$ since both represent volume per particle. Fourier transforming both sides  we get
\begin{equation}
\delta_{\bld k}(t) \equiv \int d^3\bld x   {\rm e}^{-i\bld k \cdot \bld x} \delta (\bld x,t) =   \int d^3 {\bld q} \  {\rm exp}[ - i {\bf k} . {\bf x}_{T} (t, \bld q)]  -(2 \pi)^3 \delta_D (\bld k)
\end{equation}
Differentiating this expression, 
and using  Eq.~(\ref{traj}) for the trajectories give one can obtain an equation for $\delta_{\mathbf{k}}$
(see e.g. ref.\cite{gc1}; Eq.10).
The structure of this equation can be simplified if we use the perturbed gravitational potential (in Fourier space) $\phi_{\bf k}$ related to $\delta_{\bf k}$ by 
\begin{equation}
\delta_{\bf k} = - {k^2\phi_{\bld k} \over 4 \pi G \rho_b a^2} = - \lb {k^2 a \over 4 \pi G \rho_0}\rb \phi_{\bld k} = - \lb {2 \over 3H_0^2 }\rb k^2a \phi_{\bld k}
\end{equation}
In terms of $\phi_{\bld k}$ the  \textit{exact} evolution equation reads as:
\begin{eqnarray}
\ddot \phi_{\bf k} + 4 {\dot a \over a} \dot\phi_{\bf k}   &= & - {1 \over 2a^2} \int {d^3{\bf p} \over (2 \pi )^3} \phi_{{1\over 2}{\bf k+p}}   
\phi_{{1\over 2}{\bf k-p}}\left[ \left( {k\over 2}\right)^2 + p^2 
-2  \lb {\bld k . \bld p \over k}\rb^2 \right] \nonumber \\
&+ &\lb{3H_0^2 \over 2}\rb  \int {d^3{\bf q} \over a} \lb{\bld k} . \dot {\bld x}\over k\rb ^2 e^{i{\bf k}.{\bf x}} \label{powtransf} 
\label{evphi}
\end{eqnarray}
where $\bld x = \bld x_T(t, \bld q)$. 
Of course, this equation is not `closed'.
 It contains the velocities of the particles $\dot {\bf x}_T$ and their positions explicitly in the second term on the right and one cannot --- in general --- express them in simple form in terms of $\phi_\mathbf{k}$. As a result, it might seem that we are in no better position than when we started. I will now motivate a strategy to tame this term in order to close this equation. This strategy  depends on two features: 
 
 (1) First, extremely nonlinear structures do not contribute to the right hand side of Eq.(\ref{evphi}) though, of course, they contribute individually to the two terms. 
   More precisely, the right hand side of Eq.(\ref{evphi}) will lead to a density contrast that
    will scale as 
    $\delta_{\bf k} \propto k^2$ if originally  --- in linear 
   theory --- $\delta_k \propto k^n$ with $n>2$ as $k\to 0$.  This leads to a $P\propto k^4$ tail in the power spectrum \cite{lssu,gc1}. (We will provide a derivation of this result in the next section.)

  (ii) Second, we can use Zeldovich approximation to evaluate this term, once the above fact is realized.
  It is well known that, when the density contrasts are small, it grows as
   $\delta_{\bf k}\propto a$  in the linear limit. One can easily show that such a growth corresponds to particle displacements of the form

\begin{equation}
 \bld x_{T} (a,{\bf q}) = {\bf q}  - a \nabla \psi(\mathbf{q});\qquad \psi\equiv (4\pi G\rho_0)^{-1}\phi
 \label{trajec}
\end{equation} 
A useful approximation to describe the quasi linear stages of clustering is obtained by using the trajectory in Eq.(\ref{trajec})  as an ansatz valid {\it even at quasi linear epochs}. In this approximation, (called  Zeldovich approximation), the velocities $\dot\mathbf{x}$ can be expressed in terms of the initial gravitational potential.

We now combine the two results mentioned above  to obtain a closure condition
for our dynamical equation. At any given moment of time
we can divide the particles in the system into three different sets. First, there are particles which are 
already a part of virialized cluster in the non linear regime. Second set is made of particles
which are completely unbound and are essentially contributing to power spectrum at the 
linear scales. The third set is made of all particles which cannot be put into either of these
two baskets. Of these three, we know that the first two sets  of particles do not
contribute significantly to the right hand side of Eq.(\ref{evphi}) so we will not incur any serious error in ignoring these particles in computing
the right hand side. For the description of particles 
in the third set, the Zeldovich approximation should be fairly good. In fact, we can do slightly
better than the standard Zeldovich approximation. We note that in Eq.~(\ref{trajec}) the velocities were
taken to be proportional to the gradient of the \textit{initial} gravitational potential.
We can improve on this ansatz by taking the velocities to be given by the gradient of the 
\textit{instantaneous} gravitational potential which has the effect of incorporating the
influence of particles in bound clusters on the rest of the particles to certain extent.

Given this ansatz, it is straightforward to obtain a \textit{closed} integro-differential
equation for the gravitational potential. Direct calculation shows that the gravitational potential is described by the closed integral equation: (The details of this derivation
can be found in ref. \cite{gc1} and will not be repeated here.)
\begin{equation}
\ddot \phi_{\bld k} + 4 {\dot a \over a} \dot \phi_{\bld k} = -{1 \over 3a^2} \int {d^3 \bld p \over (2 \pi)^3} \phi_{{1 \over 2} \bld k + \bld p} \phi_{{1 \over 2} \bld k - \bld p} 
\left[
{7 \over 8} k^2 + {3 \over 2} p^2 - 5 \lb {\bld k \cdot \bld p\over k}\rb^2
\right] 
\label{key1}
\end{equation}
This equation provides a powerful method for analyzing non linear clustering since estimating Eq.(\ref{evphi}) by Zeldovich approximation has a very large domain of applicability. 
 In the next two sections, I will use this equation to study the transfer of power in gravitational clustering.

 \section{\label{nltail}
 Inverse cascade in non linear gravitational clustering: The $k^4$ tail}

  There is an interesting and curious result which is characteristic of gravitational
  clustering that can be obtained directly from our Eq.~(\ref{key1}). Consider an initial
  power spectrum which has very little power at large scales; more precisely, we shall
  assume that $P(k)\propto k^n$ with $n>4$ for small $k$ (i.e, the power dies faster than $k^4$ for small $k$). If these large spatial scales are described
  by linear theory --- as one would have normally expected --- then the power at these scales 
  can only grow as $P\propto a^2k^n$ and it will always be sub dominant to $k^4$. It turns out that this 
  conclusion is incorrect.   
As the system evolves, small scale nonlinearities will
develop in the system and --- if the large scales have too little
power intrinsically  ---  then
the long wavelength power will soon be dominated by the
``$k^4$-tail'' of the short wavelength power arising from the
nonlinear clustering.  This is a purely non linear effect which we shall now describe.
(This result is known in literature \cite{lssu,gc1} but we will derive it from the formalism developed in the last section which adds fresh insight.)

\begin{figure}[ht]
\begin{center}
\includegraphics[width=.8\textwidth]{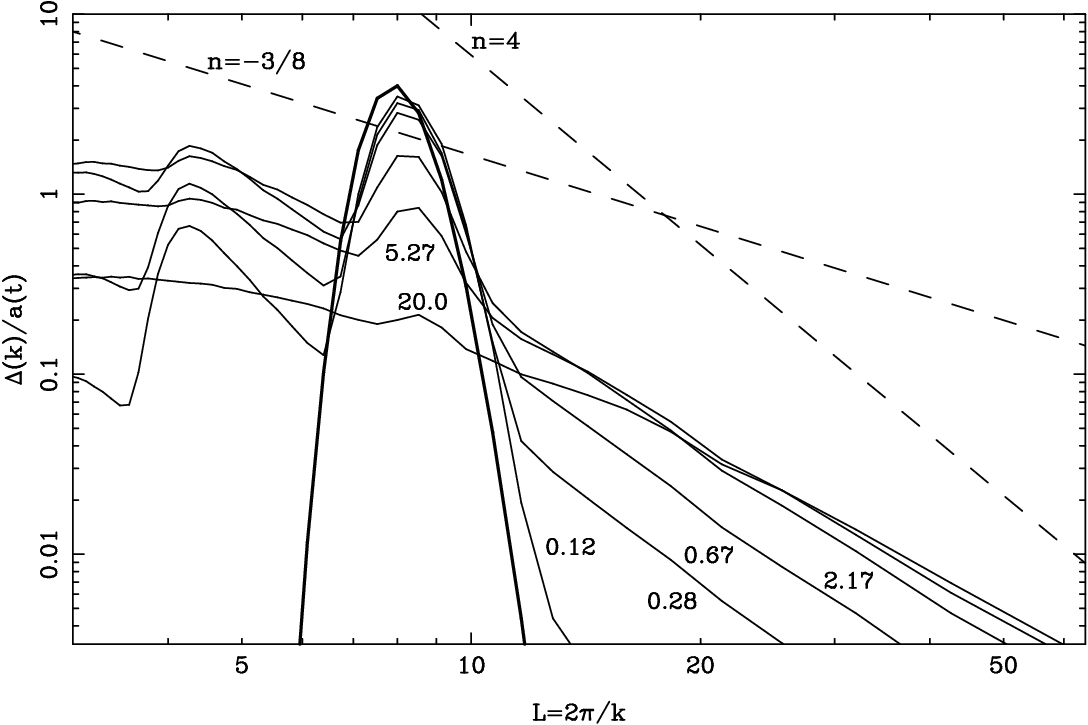}
\end{center}
\caption{The transfer of power to long wavelengths forming a $k^4$ tail is illustrated using 
simulation results. Power is injected in the form of a narrow peak at $L=8$. 
Note that the $y-$axis is $(\Delta/a)$ so that
there will be no change of shape of the power spectrum under linear evolution
with $\Delta\propto a$. As time goes on a $k^4$ tail is generated purely due to nonlinear coupling between the modes. (Figure adapted from ref.\cite{jsbtp1}.)}
\label{figptsimu}
\end{figure}
  
A  formal way of obtaining the $k^4$ tail is to solve  Eq.~(\ref{key1}) for long wavelengths; i.e. near $\bld k = 0$.   Writing $\phi_{\bld k} = \phi_{\bld k}^{(1)} + \phi_{\bld k}^{(2)} + ....$ where $\phi_{\bld k}^{(1)} = \phi_{\bld k}^{(L)}$ is the time {\it independent} gravitational potential in the linear theory and $\phi_{\bld k}^{(2)}$ is the next order correction, we get from Eq.~(\ref{key1}), the equation  
\begin{equation}
\ddot\phi_{\bld k}^{(2)}+ 4 {\dot a \over a} \dot\phi_{\bld k}^{(2)} \cong - { 1 \over 3a^2} \int {d^3 \bld p \over (2 \pi)^3} \phi^L_{{1 \over 2} \bld k + \bld p} 
\phi^L_{{1 \over 2} \bld k - \bld p} \mathcal{G}(\bld k, \bld p)
\end{equation}
where $\mathcal{G}(\bld k, \bld p)\equiv [(7/8) k^2 + (3/2) p^2 - 5(\bld k \cdot \bld p/k))^2$.
The solution to this equation is the sum of a solution to the homogeneous part [which decays as 
$\dot\phi\propto a^{-4}\propto t^{-8/3}$ giving $\phi\propto t^{-5/3}$] and a particular solution which grows as $a$. Ignoring the decaying mode at late times and taking
$\phi_{\bld k}^{(2)} = aC_{\bld k}$ one can determine $C_{\bld k}$ from the above equation. Plugging it back, we find the lowest order correction to be,
\begin{equation}
\phi_{\bld k}^{(2)} \cong - \lb {2a \over 21H^2_0}\rb \int {d^3 \bld p \over (2 \pi)^3}\phi^L_{{1 \over 2} \bld k + \bld p} 
\phi^L_{{1 \over 2} \bld k - \bld p} \mathcal{G}(\bld k, \bld p)
\label{approsol}
\end{equation}
Near $\bld k \simeq 0$, we have
\begin{eqnarray}
\phi_{\bld k \simeq 0}^{(2)} &\cong& - {2a \over 21H^2_0} \int {d^3 \bld p \over (2 \pi)^3}|\phi^L_{\bld p}|^2 \left[ {7 \over 8}k^2 + {3 \over 2}p^2 - {5(\bld k \cdot \bld p)^2 \over k^2} \right] \nonumber \\
&=&   {a \over 126 \pi^2H_0^2} \int\limits^{\infty}_0 dp  p^4 |\phi^{(L)}_{\bld p}|^2\nonumber \\
\end{eqnarray}
which is independent of $\bld k$ to the lowest order. Correspondingly the power spectrum for
 density $P_{\delta}(k)\propto a^2 k^4P_{\varphi} (k) \propto a^4 k^4$ in this order. 

The generation of long wavelength $k^4$ tail is easily seen in simulations if one starts with a power spectrum that is sharply peaked in $|\bld k|$. Figure \ref{figptsimu} (adapted from \cite{jsbtp1}) shows the results of such a simulation.   The y-axis is $[\Delta(k)/a(t)]$
where $\Delta^2(k) \equiv k^3P/2\pi^2$ is the power per logarithmic band in $k$. 
 In linear theory $\Delta \propto a$ and this quantity should not change. The curves labelled by $a=0.12$ to $a=20.0$ show the \textit{effects of nonlinear evolution}, especially the development of $k^4$ tail.
 (Actually one can do better. The formation can also reproduce the sub-harmonic at $L\simeq 4$ seen in 
 Fig.~ \ref{figptsimu} and other details; see ref. \cite{gc1}.)

\section{Analogue of Kolmogorov spectrum for gravitational clustering}

If 
power is injected at some scale $L$ into an ordinary viscous fluid, it cascades down to smaller scales because of the
non linear coupling between different modes. The resulting power spectrum, for a wide range of scales, is well approximated by the Kolmogorov spectrum which plays a key, useful, role in the study of fluid turbulence. It is possible to obtain the form of this spectrum
from fairly simple and general considerations though the actual equations of fluid turbulence are 
intractably complicated. 
 Let us now consider the corresponding question for non linear gravitational clustering. If power is injected at a given length scale
very early on, how does the dynamical evolution transfers power to other scales at late times?
In particular, does the non linear evolution lead to an analogue of Kolmogorov spectrum with some level of universality, in the case
of gravitational interactions?

Surprisingly, the answer is ``yes", even though normal fluids and collisionless self gravitating particles constitute very different physical systems. If power is injected
at a given scale $L = 2\pi/k_0$ then the gravitational clustering transfers the power to both
larger and smaller spatial scales. At large spatial scales the power spectrum goes as $P(k) \propto k^4$
as soon as non linear coupling becomes important. We have already seen this result in the previous section.  More interestingly, the cascading of power to smaller scales leads to a 
\textit{universal pattern} at late times just as in the case of fluid turbulence. This is because, Eq.(\ref{key1}) admits solutions for the  gravitational potential of the form $\phi_{\bf k}(t) = F(t)D({\bf k}) $ at late times when the initial condition is irrelevant; here
 $F(t)$ satisfies a non linear differential equation and $D({\bf k})$ satisfies an integral
equation. One can  analyze the relevant equations analytically as well as verified the conclusions
by numerical simulations. This study (the details of which can be found in ref.\cite{stp}) confirms that non linear gravitational clustering does
lead to a universal power spectrum at late times if the power is injected at a given scale
initially. 
(In cosmology there is very little  motivation to study the transfer of power by itself
and most of the numerical simulations in the past concentrated on evolving broad band initial power spectrum. So this result was missed out.)

 Our aim is to look for \textit{late time} scale free evolution of the system exploiting the 
fact  Eq.(\ref{key1})
allows self similar solutions of the form 
$\phi_{\bf k}(t) = F(t)D({\bf k}) $. Substituting this ansatz into Eq.~(\ref{key1}) we obtain
two separate equations for $F(t)$ and $D({\bf k})$. It is also convenient at this stage
to use the expansion factor $a(t) = (t/t_0)^{2/3}$ of the matter dominated universe
as the independent variable rather than the cosmic time $t$. Then simple algebra shows that
the governing equations are 
\begin{equation}
a \frac{d^2 F}{da^2} + \frac{7}{2} \frac{dF}{da} = - F^2
\label{tevl}
\end{equation}
and 
\begin{equation}
 H_0^2 D_{\bf k} =\frac{1}{3}\int \frac{d^3{\bf p}}{(2\pi)^3} D_{{1 \over 2} \bld k + \bld p} D_{{1 \over 2} \bld k - \bld p} \mathcal{G} (\bld k, \bld p)
\label{shape}
\end{equation}
Equation~(\ref{tevl}) governs the time evolution while Eq.~(\ref{shape}) governs the shape of
the power spectrum. (The  separation ansatz, of course, has the scaling freedom
$F\to \mu F, D\to (1/\mu)D$ which will change the right hand side of Eq.~(\ref{tevl}) to $-\mu F^2$ and the left hand side of Eq.~(\ref{shape})
to $\mu H_0^2 D_{\bf k}$. But, as to be expected, our results will be independent of $\mu$; so we have set it to unity). Our interest lies in analyzing the solutions of Eq.~(\ref{tevl}) subject to the initial conditions $F(a_i) =$ constant, $(dF/da)_i =0$ at some small enough $a=a_i$.

\begin{figure*}[t]
\begin{center}
\includegraphics[scale=0.5]{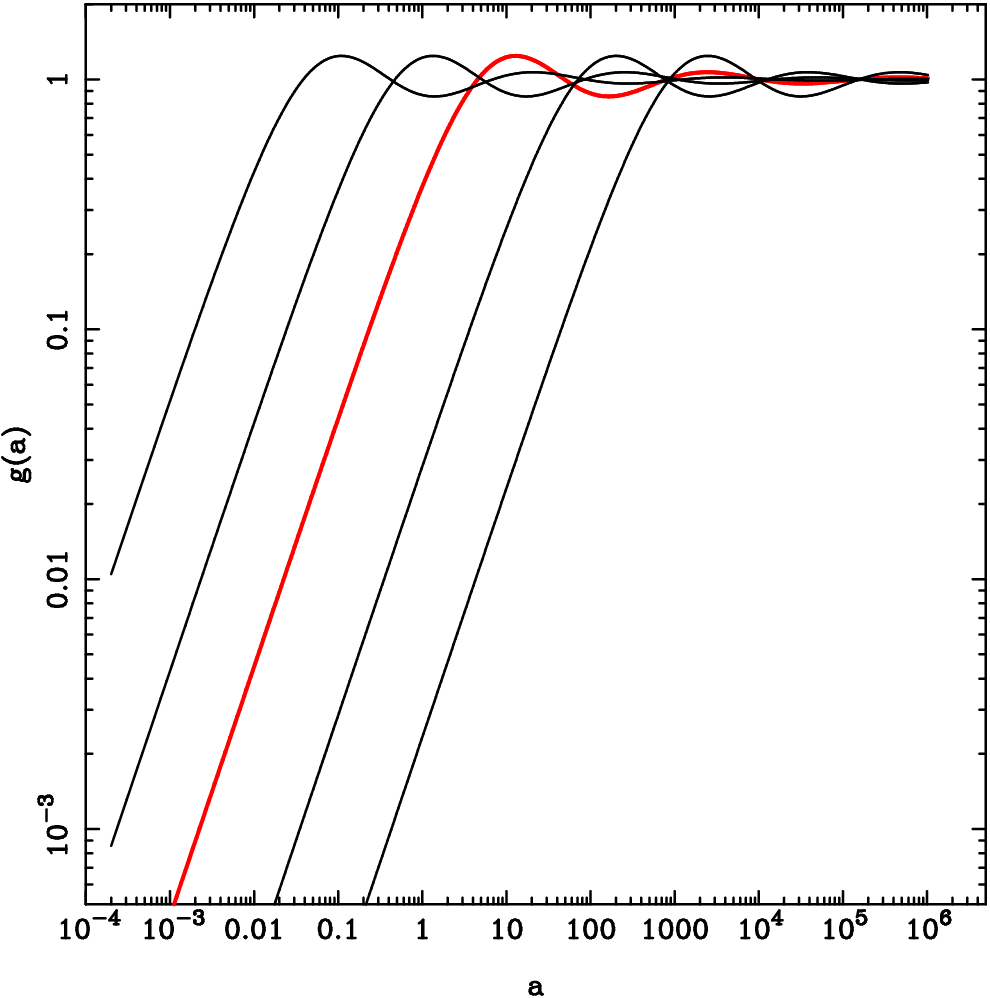}
\end{center}
\caption{The solution to Eq.~(\ref{gofa}) plotted in  $g - a$ plane. The 
function $g(a)$ asymptotically approaches unity with oscillations which are represented by the
spiral in the right panel. The different curves in the  left panel corresponds to the rescaling freedom in the 
initial conditions. One fiducial curve which was used to model the simulation is shown in the
red. For more details, see ref.\cite{stp}.}
\label{gc1}
\end{figure*}

\begin{figure*}
\includegraphics[scale=0.4]{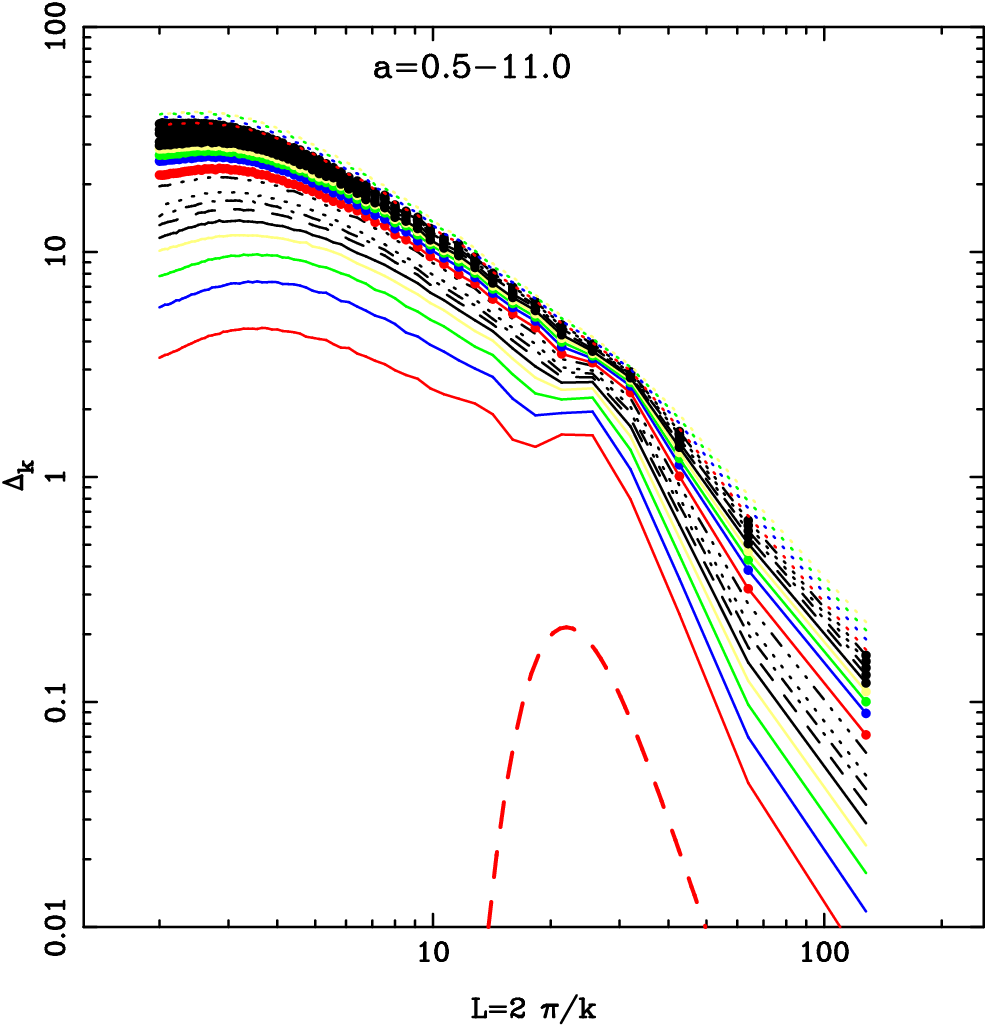}\ \includegraphics[scale=0.4]{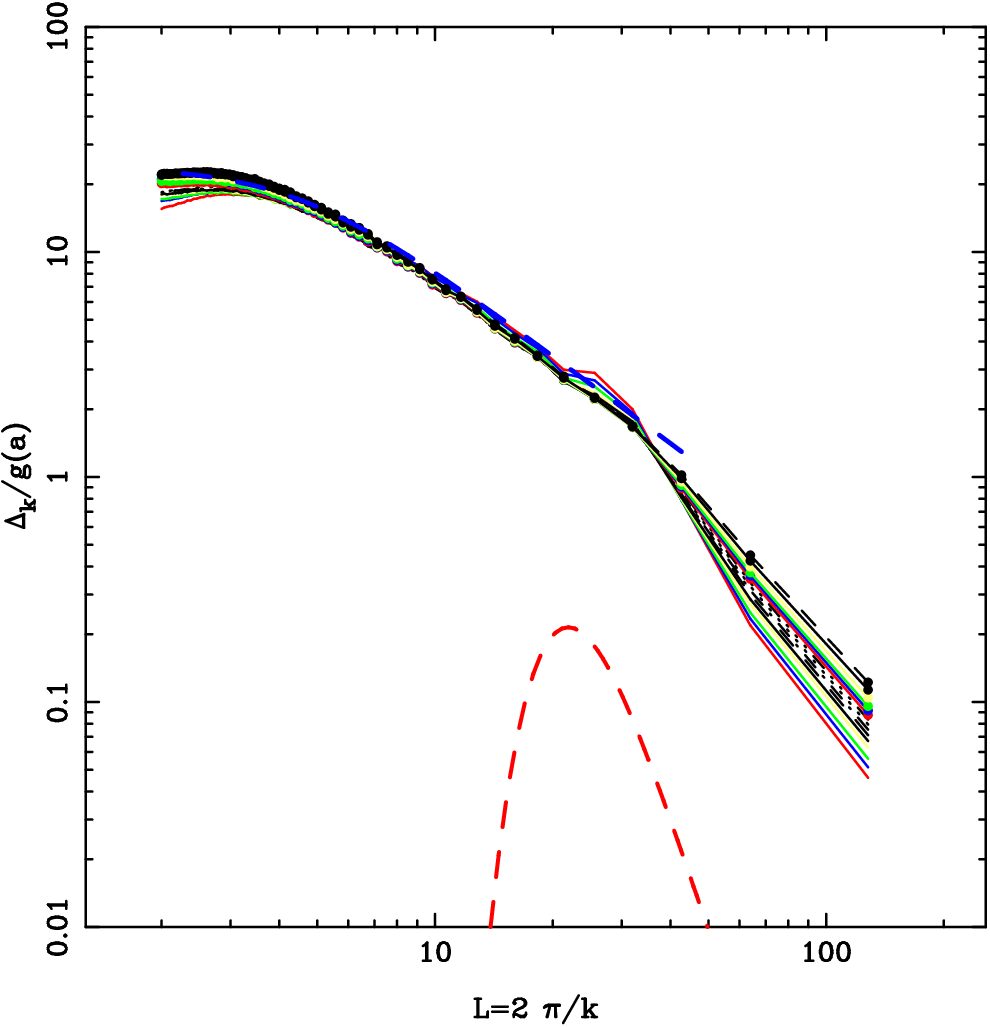}
\caption{Left panel: The results of the numerical simulation with an initial power spectrum which is 
a Gaussian peaked at $L=24$. The y-axis gives $\Delta_k$ where $\Delta_k^2=k^3 P/2\pi^2$ is
the power per logarithmic band. The evolution generates a well known $k^4$ tail at large scales (see for example, \cite {jsbtp1})
and leads to cascading of power to small scales. Right panel: The simulation data is re-expressed
by factoring out the time evolution function $g(a)$ obtained by integrating Eq.~(\ref{gofa}). The fact
that the curves fall nearly on top of each other shows that the late time evolution is scale free
and described by the ansatz discussed in the text. The rescaled spectrum is very well described by
$P(k)\propto k^{-0.4}/(1+(k/k_0)^{2.6})$ which is shown by the, completely overlapping, broken blue curve.
For more details, see ref.\cite{stp}.}
\label{fig:gausspeak}
\end{figure*}

Inspection shows that Eq.~(\ref{tevl}) has the exact solution $F(a) = (3/2) a^{-1}$.
This, of course, is a special solution and will not satisfy the relevant initial conditions. However,
Eq.~(\ref{tevl}) fortunately belongs to a  class of non linear equations which can be mapped
to a homologous system. In such cases, the special power law solutions will arise as
the asymptotic limit. (The example well known to astronomers is that of isothermal sphere \cite{chandra}. Our
analysis below has a close parallel.)
To find the general behaviour of the solutions to Eq.~(\ref{tevl}), we will make the substitution
$F(a) = 
(3/2) a^{-1} g(a)$ and change the independent variable from $a$ to $q = \log a$.
Then Eq.~(\ref{tevl}) reduces to the form
 \begin{equation}
\frac{d^2 g}{d q^2} + \frac{1}{2} \frac{d g}{d q} + \frac{3}{2} 
g(g - 1) = 0
\label{gofa}
\end{equation} 
This represents a particle moving in a potential $V(g)= (1/2) g^3 - (3/4) g^2$ under 
friction.  For our initial conditions the motion will lead the ``particle'' to asymptotically come to rest at the stable minimum
at $g=1$ with damped oscillations. In other words, $F(a) \to (3/2) a^{-1}$ for large $a$
showing this is indeed the asymptotic solution. From the Poisson equation, it follows that $k^2\phi_\mathbf{k}\propto (\delta_\mathbf{k}/a)$ so that $\delta_\mathbf{k}(a)\propto g(a)k^2D(\mathbf{k})$ giving a direct physical meaning to the function $g(a)$ as the growth factor for the density contrast. The asymptotic limit ($g\simeq 1$) corresponds to to a rather trivial case of $\delta_\mathbf{k}$ becoming independent of time. What will be more interesting --- and accessible in simulations --- will be the approach to this asymptotic solution. To obtain this, we introduce
the variable 
\begin{equation}
u = g + 2 \left( \frac{dg}{dq}\right)
\end{equation}
so that our system becomes homologous. It can be easily shown that 
we now get the first order form of the autonomous system to be 
\begin{equation} 
\frac{d u}{d g} = - \frac{6 g(g - 1)} {u - g}
\end{equation}
The critical points  of the system are at $(0,0) $ and $(1,1)$. Standard analysis shows that: (i) the  ($0,0$) is an unstable critical point and the second one $(1,1)$ is the stable critical point; (ii) for our initial
conditions the solution spirals around the stable critical point.

Figure \ref{gc1} (from ref. \cite{stp})  describes the solution in the $g-a$  plane. The $g(a)$ curves clearly approach the asymptotic 
value of $g\approx 1$ with superposed oscillations. The different curves in Fig. (\ref{gc1})
are for different initial values which arise from the scaling freedom mentioned earlier. (The thick red line correspond to the initial conditions used in the simulations described below.).  The solution
$g(a)$ describes the time evolution and solves the problem of determining asymptotic time evolution.

To test the correctness of these conclusions, we performed a high resolution simulation 
using the TreePM method
\cite{2002JApA...23..185B,2003NewA....8..665B} and its parallel version
\cite{2004astro.ph..5220Ra} with $128^3$ particles
on a $128^3$ grid. 
Details about the code parameters can be found in
\cite{2003NewA....8..665B}.
The initial power spectrum $P(k)$ was chosen to be a Gaussian peaked at
the scale of $k_p = 2 \pi / L_p$ with $L_p = 24$ grid lengths and with a
standard deviation $\Delta k = 2 \pi / L_{box}$, where $L_{box} = 128$ is
the size of one side of the simulation volume. The amplitude of the peak
was taken such that $\Delta_{lin}\left(k_p = 2 \pi / L_p, a = 0.25\right)
= 1$.
 
The  late time evolution of the power spectrum (in terms of $\Delta_k^2\equiv k^3P(k)/2\pi^2$ where $P=|\delta_k|^2$ is the power spectrum of density fluctuations) obtained from the simulations is shown in Fig.\ref{fig:gausspeak} (left panel). In the right panel, we have rescaled the $\Delta_k$, using the appropriate solution $g(a)$. The fact that the curves fall on top of each other shows that the late time evolution indeed sales as $g(a)$ within numerical accuracy.  A reasonably accurate fit for $g(a)$ at late times used in this figure is given by $
g(a)\propto a(1-0.3\ln a)$.
The key point to note is that the asymptotic time evolution is essentially $\delta(a)\propto a$ except for a logarithmic correction, \textit{even in highly nonlinear scales}. (This was first noticed from somewhat lower resolution simulations in \cite{jsbtp1}.). Since the evolution at \textit{linear} scales is always $\delta\propto a$,
 this allows for a form invariant evolution of power spectrum at all scales. Gravitational clustering evolves towards this asymptotic state. 
 
 To the lowest order of accuracy, the power spectrum at this range of scales is approximated by the mean index $n\approx -1$ with $P(k)\propto k^{-1}$. A better fit to the 
  power spectrum in Fig.\ref{fig:gausspeak} is given by
\begin{equation}
P(k)\propto \frac{k^{-0.4}}{1+(k/k_0)^{2.6}}; \qquad \frac{2\pi}{k_0}\approx 4.5
\label{fit}
\end{equation}
This fit is shown by the broken blue line in the figure which completely overlaps with the data and is barely visible. (Note that this fit is applicable only at $L<L_p$ since the $k^4$ tail will dominate scales to the right of the initial peak; see the discussion in \cite {jsbtp1}). At nonlinear scales $P(k)\propto k^{-3}$
making $\Delta_k$ flat, as seen in Fig. \ref{fig:gausspeak}. (This is \textit{not} a numerical artifact and we have sufficient dynamic range in the simulation to ascertain this.) At quasi linear scales
$P(k)\propto k^{-0.4}$. The effective index of the power spectrum varies between $-3$ and $-0.4$ in this range of scales. 

\section{Possible interpretation}

In the case of viscous fluids, the energy is dissipated at the smallest scales as heat. In steady state, energy cannot accumulate at any intermediate scale and hence the rate of flow of energy from one scale to the next (lower) scale must be a constant. This constancy immediately leads to Kolmogorov spectrum. In the case of gravitating particles, there is no dissipation and each scale will evolve towards virial equilibrium. At any given time $t$, the power would have cascaded down only up to some scale $l_{min}(t)$ which it self, of course, is a decreasing function of  time. So, at time $t$ we expect very little power for $1<kl_{min}(t)$ and a $k^4$ tail for $kL_{p}<1$, say. The really interesting band is between $l_{min}$ and $L_{p}$. 

To understand this band, let us recall that the Lagrangian in Eq.(\ref{basicL}) leads to the time dependent Hamiltonian is $H(\mathbf{p},\mathbf{x},t)=\sum[p^2/2ma^2+U]$. The evolution of the energy in the system is governed by the equation $dH/da=(\partial H/\partial a)_{\mathbf{p},\mathbf{x}}.$
It is clear from Eq.(\ref{defphi}) that $(\partial U/\partial a)_{\mathbf{p},\mathbf{x}}=-U/a$ while $(\partial T/\partial a)_{\mathbf{p},\mathbf{x}}=-2T/a$. Hence the time evolution of the total energy $H=E$ of the system is described by
\begin{equation}
\frac{dE}{da}=-\frac{1}{a}(2T+U)=-\frac{2E}{a}+\frac{U}{a}=-\frac{E}{a}-\frac{T}{a}
\end{equation} 
In the continuum limit, ignoring the infinite self-energy term,  the potential energy can be written as:
\begin{equation}
U=-\frac{G\rho_0^2}{2a}\int d^3\mathbf{x}\int d^3\mathbf{y}
\frac{\delta(\mathbf{x},a)\delta(\mathbf{y},a)}{|\mathbf{x}-\mathbf{y}|}
\end{equation} 
Hence
\begin{equation}
\frac{d (a^2E)}{da}=aU=-\frac{G\rho_0^2}{2}\int d^3\mathbf{x}\int d^3\mathbf{y}
\frac{\delta(\mathbf{x},a)\delta(\mathbf{y},a)}{|\mathbf{x}-\mathbf{y}|}
\label{pottot}
\end{equation} 
The ensemble average of $U$, per unit  volume, will be
\begin{equation}
\mathcal{E}= -\frac{G\rho_0^2}{2Va}\int d^3\mathbf{x} d^3\mathbf{y}
\frac{\langle \delta(\mathbf{x},a)\delta(\mathbf{y},a)\rangle}{|\mathbf{x}-\mathbf{y}|}
\propto \int d^3k \frac{|\delta_k|^2}{k^2} 
\propto\int_0^\infty \frac{dk}{k} kP(k)
\label{enden}
\end{equation}
where $V$ is the comoving volume.

When a particular scale is virialized, we expect $\mathcal{E}\approx$ constant at that scale in comoving coordinates. That is, we would expect
the energy density in Eq.(\ref{enden}) would have  reached equipartition and contribute same amount per logarithmic band of scales
 in the intermediate scales between $l_{min}$ and $L_{peak}$. 
\textit{ This requires $P(k)\propto 1/k$ which is essentially what we found from simulations.}
The time dependence of $P$ is essentially $P\propto a^2$ (except for a logarithmic correction) in order to maintain form invariance of the spectrum with respect to the linear end. Similarly the scale dependence is $P\propto k^{-1}$ 
 which is indeed a good fit to the simulation results. The flattening of the power at small scales, modeled by the more precise fitting function in Eq.(\ref{fit}), can be understood from the fact that, equipartition is not yet achieved at smaller scales. The same result holds for kinetic energy if the motion is dominated by scale invariant radial flows \cite{jsbtp1,klypin}. Our result suggests that gravitational power transfer evolves towards this equipartition. 

\section*{Acknowledgements}

Part of the work reported here was done in collaboration with Suryadeep Ray and J.S.Bagla. Numerical experiments for this study were carried out at the cluster
computing facility in the Harish-Chandra Research Institute.

 \end{document}